\documentclass[aps,prl,superscriptaddress,showpacs,floatfix,twocolumn]{revtex4}
\usepackage{graphicx}	% Include figure files

\def\Journal#1#2#3#4{{#1}{\bf #2}, #3 (#4)}

\def\NIMA{{Nucl. Instr. Meth.}~{\bf A}}

\def\PLB{{Phys. Lett.}~{\bf B}}
\def\PLC{Phys. Repts.\ }
\def\PRL{Phys. Rev. Lett.\ }
\def\PRD{{Phys. Rev.}~{\bf D}}

\begin{document}

\title{$J/\psi$ Production vs Transverse Momentum and Rapidity \\
in $p$+$p$ Collisions at $\sqrt{s}= 200$~GeV
}

\newcommand{\abilene}{Abilene Christian University, Abilene, TX 79699, U.S.}
\newcommand{\banaras}{Department of Physics, Banaras Hindu University, Varanasi 221005, India}
\newcommand{\bnl}{Brookhaven National Laboratory, Upton, NY 11973-5000, U.S.}
\newcommand{\caucr}{University of California - Riverside, Riverside, CA 92521, U.S.}
\newcommand{\charlesczech}{Charles University, Ovocn\'{y} trh 5, Praha 1, 116 36, Prague, Czech Republic}
\newcommand{\ciae}{China Institute of Atomic Energy (CIAE), Beijing, People's Republic of China}
\newcommand{\cns}{Center for Nuclear Study, Graduate School of Science, University of Tokyo, 7-3-1 Hongo, Bunkyo, Tokyo 113-0033, Japan}
\newcommand{\colorado}{University of Colorado, Boulder, CO 80309, U.S.}
\newcommand{\columbia}{Columbia University, New York, NY 10027 and Nevis Laboratories, Irvington, NY 10533, U.S.}
\newcommand{\czechtech}{Czech Technical University, Zikova 4, 166 36 Prague 6, Czech Republic}
\newcommand{\dapnia}{Dapnia, CEA Saclay, F-91191, Gif-sur-Yvette, France}
\newcommand{\debrecen}{Debrecen University, H-4010 Debrecen, Egyetem t{\'e}r 1, Hungary}
\newcommand{\elte}{ELTE, E{\"o}tv{\"o}s Lor{\'a}nd University, H - 1117 Budapest, P{\'a}zm{\'a}ny P. s. 1/A, Hungary}
\newcommand{\fit}{Florida Institute of Technology, Melbourne, FL 32901, U.S.}
\newcommand{\fsu}{Florida State University, Tallahassee, FL 32306, U.S.}
\newcommand{\gsu}{Georgia State University, Atlanta, GA 30303, U.S.}
\newcommand{\hiroshima}{Hiroshima University, Kagamiyama, Higashi-Hiroshima 739-8526, Japan}
\newcommand{\ihepprot}{IHEP Protvino, State Research Center of Russian Federation, Institute for High Energy Physics, Protvino, 142281, Russia}
\newcommand{\illuiuc}{University of Illinois at Urbana-Champaign, Urbana, IL 61801, U.S.}
\newcommand{\instpasczech}{Institute of Physics, Academy of Sciences of the Czech Republic, Na Slovance 2, 182 21 Prague 8, Czech Republic}
\newcommand{\isu}{Iowa State University, Ames, IA 50011, U.S.}
\newcommand{\jinrdubna}{Joint Institute for Nuclear Research, 141980 Dubna, Moscow Region, Russia}
\newcommand{\kek}{KEK, High Energy Accelerator Research Organization, Tsukuba, Ibaraki 305-0801, Japan}
\newcommand{\kfki}{KFKI Research Institute for Particle and Nuclear Physics of the Hungarian Academy of Sciences (MTA KFKI RMKI), H-1525 Budapest 114, POBox 49, Budapest, Hungary}
\newcommand{\korea}{Korea University, Seoul, 136-701, Korea}
\newcommand{\kurchatov}{Russian Research Center ``Kurchatov Institute", Moscow, Russia}
\newcommand{\kyoto}{Kyoto University, Kyoto 606-8502, Japan}
\newcommand{\labllr}{Laboratoire Leprince-Ringuet, Ecole Polytechnique, CNRS-IN2P3, Route de Saclay, F-91128, Palaiseau, France}
\newcommand{\lawllnl}{Lawrence Livermore National Laboratory, Livermore, CA 94550, U.S.}
\newcommand{\losalamos}{Los Alamos National Laboratory, Los Alamos, NM 87545, U.S.}
\newcommand{\lpc}{LPC, Universit{\'e} Blaise Pascal, CNRS-IN2P3, Clermont-Fd, 63177 Aubiere Cedex, France}
\newcommand{\lund}{Department of Physics, Lund University, Box 118, SE-221 00 Lund, Sweden}
\newcommand{\muenster}{Institut f\"ur Kernphysik, University of Muenster, D-48149 Muenster, Germany}
\newcommand{\myongji}{Myongji University, Yongin, Kyonggido 449-728, Korea}
\newcommand{\nagasaki}{Nagasaki Institute of Applied Science, Nagasaki-shi, Nagasaki 851-0193, Japan}
\newcommand{\newmex}{University of New Mexico, Albuquerque, NM 87131, U.S. }
\newcommand{\nmsu}{New Mexico State University, Las Cruces, NM 88003, U.S.}
\newcommand{\ornl}{Oak Ridge National Laboratory, Oak Ridge, TN 37831, U.S.}
\newcommand{\orsay}{IPN-Orsay, Universite Paris Sud, CNRS-IN2P3, BP1, F-91406, Orsay, France}
\newcommand{\peking}{Peking University, Beijing, People's Republic of China}
\newcommand{\pnpi}{PNPI, Petersburg Nuclear Physics Institute, Gatchina, Leningrad region, 188300, Russia}
\newcommand{\riken}{RIKEN, The Institute of Physical and Chemical Research, Wako, Saitama 351-0198, Japan}
\newcommand{\rikjrbrc}{RIKEN BNL Research Center, Brookhaven National Laboratory, Upton, NY 11973-5000, U.S.}
\newcommand{\rikkyo}{Physics Department, Rikkyo University, 3-34-1 Nishi-Ikebukuro, Toshima, Tokyo 171-8501, Japan}
\newcommand{\saispbstu}{Saint Petersburg State Polytechnic University, St. Petersburg, Russia}
\newcommand{\saopaulo}{Universidade de S{\~a}o Paulo, Instituto de F\'{\i}sica, Caixa Postal 66318, S{\~a}o Paulo CEP05315-970, Brazil}
\newcommand{\seoulnat}{System Electronics Laboratory, Seoul National University, Seoul, South Korea}
\newcommand{\stonybrkc}{Chemistry Department, Stony Brook University, Stony Brook, SUNY, NY 11794-3400, U.S.}
\newcommand{\stonycrkp}{Department of Physics and Astronomy, Stony Brook University, SUNY, Stony Brook, NY 11794, U.S.}
\newcommand{\subatech}{SUBATECH (Ecole des Mines de Nantes, CNRS-IN2P3, Universit{\'e} de Nantes) BP 20722 - 44307, Nantes, France}
\newcommand{\tenn}{University of Tennessee, Knoxville, TN 37996, U.S.}
\newcommand{\titech}{Department of Physics, Tokyo Institute of Technology, Oh-okayama, Meguro, Tokyo 152-8551, Japan}
\newcommand{\tsukuba}{Institute of Physics, University of Tsukuba, Tsukuba, Ibaraki 305, Japan}
\newcommand{\vandy}{Vanderbilt University, Nashville, TN 37235, U.S.}
\newcommand{\waseda}{Waseda University, Advanced Research Institute for Science and Engineering, 17 Kikui-cho, Shinjuku-ku, Tokyo 162-0044, Japan}
\newcommand{\weizmann}{Weizmann Institute, Rehovot 76100, Israel}
\newcommand{\yonsei}{Yonsei University, IPAP, Seoul 120-749, Korea}
\affiliation{\abilene}
\affiliation{\banaras}
\affiliation{\bnl}
\affiliation{\caucr}
\affiliation{\charlesczech}
\affiliation{\ciae}
\affiliation{\cns}
\affiliation{\colorado}
\affiliation{\columbia}
\affiliation{\czechtech}
\affiliation{\dapnia}
\affiliation{\debrecen}
\affiliation{\elte}
\affiliation{\fit}
\affiliation{\fsu}
\affiliation{\gsu}
\affiliation{\hiroshima}
\affiliation{\ihepprot}
\affiliation{\illuiuc}
\affiliation{\instpasczech}
\affiliation{\isu}
\affiliation{\jinrdubna}
\affiliation{\kek}
\affiliation{\kfki}
\affiliation{\korea}
\affiliation{\kurchatov}
\affiliation{\kyoto}
\affiliation{\labllr}
\affiliation{\lawllnl}
\affiliation{\losalamos}
\affiliation{\lpc}
\affiliation{\lund}
\affiliation{\muenster}
\affiliation{\myongji}
\affiliation{\nagasaki}
\affiliation{\newmex}
\affiliation{\nmsu}
\affiliation{\ornl}
\affiliation{\orsay}
\affiliation{\peking}
\affiliation{\pnpi}
\affiliation{\riken}
\affiliation{\rikjrbrc}
\affiliation{\rikkyo}
\affiliation{\saispbstu}
\affiliation{\saopaulo}
\affiliation{\seoulnat}
\affiliation{\stonybrkc}
\affiliation{\stonycrkp}
\affiliation{\subatech}
\affiliation{\tenn}
\affiliation{\titech}
\affiliation{\tsukuba}
\affiliation{\vandy}
\affiliation{\waseda}
\affiliation{\weizmann}
\affiliation{\yonsei}
\author{A.~Adare}	\affiliation{\colorado}
\author{S.~Afanasiev}	\affiliation{\jinrdubna}
\author{C.~Aidala}	\affiliation{\columbia}
\author{N.N.~Ajitanand}	\affiliation{\stonybrkc}
\author{Y.~Akiba}	\affiliation{\riken} \affiliation{\rikjrbrc}
\author{H.~Al-Bataineh}	\affiliation{\nmsu}
\author{J.~Alexander}	\affiliation{\stonybrkc}
\author{K.~Aoki}	\affiliation{\kyoto} \affiliation{\riken}
\author{L.~Aphecetche}	\affiliation{\subatech}
\author{R.~Armendariz}	\affiliation{\nmsu}
% ...
\collaboration{PHENIX Collaboration} \noaffiliation

\date{\today}

%%% Abstract %%%%%%%%%%%%%%%%%%%%%%%%%%%%%%%%%%%%%%

\begin{abstract}
$J/\psi$ production in $p{+}p$ collisions at $\sqrt{s}= 200$~GeV has 
been measured in the PHENIX experiment at the Relativistic Heavy Ion 
Collider (RHIC) over a rapidity range of $-2.2<y<2.2$ and a transverse 
momentum range of $0< p_{T} < 9$ GeV/$c$.
The statistics available allow a detailed measurement of both the $p_{T}$ 
and rapidity distributions and are sufficient to constrain production models.   
The total cross section times branching ratio determined for $J/\psi$ production is 
$B_{ll} \cdot \sigma_{pp}^{J/\psi} = 
178{\pm}3^{\rm stat}{\pm}53^{\rm sys}{\pm}18^{\rm norm}$~nb.
\end{abstract}

\pacs{PACS numbers: }

% It is optional to also add (uncomment):
% \keywords{}

\maketitle

% **************   put body of paper here    *******************

%01_PhysicsMotivation.tex
$J/\psi$ are produced in hadronic collisions involving hard processes 
that proceed primarily through diagrams involving gluons, such as 
gluon-gluon fusion.
Once the $c\bar{c}$ pair is produced it must evolve through a
hadronization process to form a physical $J/\psi$. While this production has
been extensively studied, the details of the production mechanism and
hadronization remain an open question. Attempts at a consistent theoretical 
description of $J/\psi$ production have been made, but it 
has proven difficult to reproduce both the observed cross sections and 
polarization~\cite{beneke,CDF,fixed_target,E789}.
An additional complication is that nearly $30$-$40\%$ 
of the measured $J/\psi$ yield results from feeddown of higher mass states 
($\psi$', $\chi_c$)~\cite{HERAB}, reducing the observed polarization with
respect to that expected from directly produced $J/\psi$.

The color-singlet model~\cite{CSM}, which generates a color singlet
$c\bar{c}$ pair in the same quantum state as the $J/\psi$,
underpredicts the measured $J/\psi$ cross section by
approximately an order of magnitude~\cite{CDF}. Alternatively, the
color-octet model includes color octet $c\bar{c}$ pairs
that radiate soft gluons during $J/\psi$ formation~\cite{COM}.  
However, the color
octet matrix elements are not universal~\cite{octet_matrix_elements}
and the predicted transverse $J/\psi$ polarization at high $p_{\rm T}$ is 
not seen in the data~\cite{CDF,E789}. 
The color evaporation model, a more phenomenological approach,
forms the different charmonium states in proportions determined from
experimental data for any $c\bar{c}$ pair that has a mass below the
$D\bar{D}$ threshold and predicts no polarization.
Finally, a recent perturbative QCD calculation 
including 3-gluon diagrams is able to successfully reproduce
both the observed cross section and polarization results~\cite{Khoze}.

A fundamental understanding of the $J/\psi$ production process is also 
critical to defining the configuration of the produced $c\bar{c}$ state 
since this will have direct implications on the interaction of this state 
with both cold nuclear
matter in proton or deuteron-nucleus collisions and with the high-density
partonic matter observed in high-energy heavy-ion collisions.
High quality experimental results over wide kinematical and collision 
energy ranges are required to constrain models and to provide an improved
understanding of $J/\psi$ (and other heavy quarkonia) production.

In this paper, $J/\psi$ production in $p{+}p$ collisions at 
$\sqrt{s}= 200$~GeV measured by the PHENIX experiment at RHIC is reported.   
The $J/\psi$ cross section and transverse momentum distributions are studied 
in the mid ($|y| \leq0.35$) and forward ($1.2 < |y| <2.2$) rapidity regions.   
The data presented were collected during the 2005 RHIC run and exceed by more 
than one order of magnitude the previously reported number of 
$J/\psi$~\cite{ppg017,ppg038}.

A detailed description of the PHENIX experiment is provided 
in~\cite{phenix_nim}. 
At mid-rapidity the drift chambers (DC), ring imaging
\v{C}erenkov detectors (RICH), and electromagnetic calorimeters
(EMCal) are used to detect $J/\psi \rightarrow e^{+}e^{-}$ decays
in two arms each covering $\Delta\phi=90^{\circ}$ in azimuth.
The muon detectors, consisting of cathode strip tracking
chambers in a magnetic field (MuTr) and alternating layers of steel 
absorber and Iarocci tube planes (MuID), are used to measure 
$J/\psi \rightarrow \mu^{+}\mu^{-}$ at forward and backward rapidities over
$\Delta\phi=360^{\circ}$.

The data were recorded using a minimum bias trigger that requires
at least one hit in each of the beam-beam counters (BBC)
at forward and backward rapidity, $3.0 < |\eta| < 3.9$.  
Di-electron events must pass an additional trigger that consisted of
an OR between the level-1 electron and photon triggers.   
The electron trigger requires matching hits between the EMCal and RICH 
in a small angular area with a minimum energy deposition of 0.4~GeV in 
any $2\times2$ patch of EMCal towers.
The photon trigger requires a minimum energy deposition of 1.4~GeV in 
any $4\times4$ set of overlapping EMCal towers.
A $J/\psi$ trigger efficiency of $96\%$ was achieved within the vertex 
range $|z_{vtx}| < 30$~cm.
Di-muon triggered events were selected using an online level-1 trigger 
that requires at least two particles penetrate the MuID.  
One particle must penetrate the entire MuID while the second has a 
minimum penetration depth of 3 out of the 5 pairs of detector and absorber 
planes.   
Approximately $92\%$ of the $J/\psi$'s within $|z_{vtx}| < 30$~cm fulfill 
this requirement. 
As part of the reconstruction chain a level-2 filter, which 
consists of a fast reconstruction of the particle trajectory in the MuTr 
and MuID is applied.  
Events are accepted by this filter when at least two particles penetrate 
the entire MuID and have a reconstructed invariant mass $\geq2.0$~GeV/$c^2$.   
After applying cuts on the collision vertex position and 
quality assurance criteria, the sampled statistics corresponds to 
$2.6~$pb$^{-1}$ in the dielectron analysis, $2.7~$pb$^{-1}$
in the muon arm covering $1.2 < y <2.2$ and $3.5~$pb$^{-1}$ in the 
muon arm covering $-2.2 < y <-1.2$.

\begin{figure}[tb]
\includegraphics[width=1.0\linewidth]{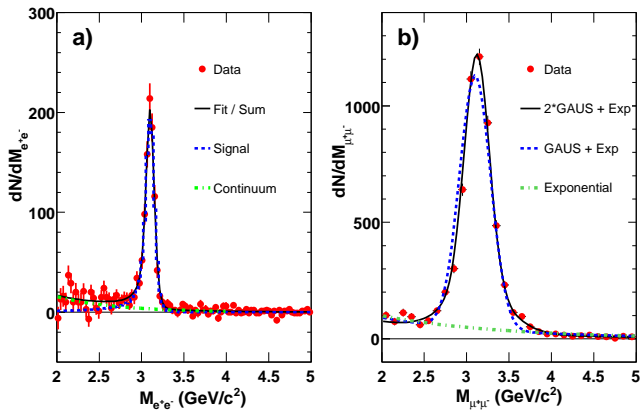}
\caption[]{Invariant mass spectra for a)$J/\psi \rightarrow e^{+} e^{-}$
at $|y| <0.35$ and b) $J/\psi \rightarrow \mu^{+} \mu^{-}$ at
$1.2 < |y| <2.2$ with the functional forms used to extract the number
of reconstructed $J/\psi$. \label{fig:invmass}}
\end{figure}

At mid-rapidity electron candidates are charged tracks associated
with at least two hit phototubes in the RICH and one EMCal hit with 
a position matching of $\pm 4$~standard deviations ($\sigma$).  
The energy-momentum matching requirement is 
\mbox{$(E/p-1)/\geq$-4.0$\sigma$}.  
The number of $J/\psi$ is obtained by counting the unlike-sign dielectron 
pairs in a fixed mass window after subtracting the like-sign pairs.
Figure~\ref{fig:invmass}(a) shows the invariant mass spectrum for 
dielectron 
pairs after subtracting the like-sign background. 
The mass window for counting the $J/\psi$ signal is 2.7~-~3.5~GeV/$c^2$ or 
2.6~-~3.6~GeV/$c^2$ depending on the number of the DC hits used to 
reconstruct the track. 
The $J/\psi$ counts are corrected for the continuum yield, which originates 
primarily from open charm pairs and Drell-Yan, inside the mass window 
($10\%\pm5\%$) and the fraction of $J/\psi$ outside of the mass 
window ($7.2\%\pm1.0\%$).
Approximately 1500 $J/\psi \rightarrow e^+e^-$ are obtained.
The solid line in the figure is the sum of the $J/\psi$ line shape 
(dashed curve) and an exponential function (dot-dashed curve) describing 
the continuum component. 
The $J/\psi$ line shape function includes the detector resolution,
the internal radiative effect~\cite{radiative_effect}, and the external 
radiative effect evaluated using a GEANT~\cite{GEANT} simulation of 
PHENIX detector.

%%%%%%%%%%%%%%%%%%%%%%%%%%%%%%%%%%%%%%%%%%%%% Fig. 1

Muon track candidates are selected based upon their penetration depth in the 
MuID and the reconstructed track quality within the MuID and MuTr. 
The particle trajectory must contain at least 8 of 10 possible hits in the 
MuID and the position matching between the MuID and MuTr must be within 
$15 (20)$~cm at positive (negative) rapidity.  
The $J/\psi$ yield is obtained from the unlike sign dimuon invariant mass
distribution by subtracting the combinatorial background estimated using an 
event mixing technique.  
Three functions, shown in Fig.~\ref{fig:invmass}(b), are used to extract 
the $J/\psi$ yield. Single Gaussian + exponential and 
double Gaussian + exponential functions are used to fit the $J/\psi$ peak, 
while the contribution from the physical continuum and background is 
estimated using an exponential fit.  
The reported number of $J/\psi$ represents the average of the fit values. 
A total of 8000 $J/\psi \rightarrow \mu^+\mu^-$ are obtained.  
The systematic error associated with the signal extraction is 
estimated from the variation between the fits.

The $J/\psi$ cross section in a given rapidity and transverse momentum bin is
calculated according to, 
\begin{displaymath}
\frac{B_{ll}}{2\pi p_{\rm T}}
\frac{ d^2\sigma_{J/\psi} }{ dy dp_{\rm T}} \,=\,
\frac{ 1 }{{ {2\pi} \, {p_{\rm T}} \, {\Delta}p_{\rm T} \,\Delta}y } \,
\frac{ N_{J/\psi} }{ {\cal L} \, A \, {\epsilon_{\rm rec}} \, {\epsilon_{\rm trig}} \, { \epsilon_{J/\psi}^{\rm BBC} } } ,
\label{eq:dsigdpt_alt}
\end{displaymath}
where $B_{ll}$ is the $J/\psi$ dilepton branching ratio, $N_{J/\psi}$ is the
measured $J/\psi$ yield,  ${\cal L}$ is the integrated luminosity recorded by
the minimum bias trigger,  $A \, \epsilon_{\rm rec}$ represents the geometrical
acceptance and reconstruction efficiency, and $\epsilon_{\rm trig}$ is the trigger
efficiency.   $\epsilon_{J/\psi}^{\rm BBC}$ is the minimum bias trigger
efficiency for events containing a $J/\psi$ and was determined to be
$0.79~\pm~0.02$~\cite{ppg017}. The cross section sampled by the BBC trigger, 
$\sigma_{\rm tot}^{pp}\times\epsilon_{\rm MB}^{\rm BBC}~=~ 23.0\pm2.2$~mb, was used to determine 
the integrated luminosity.

The $A\epsilon_{\rm rec}$ and $\epsilon_{\rm trig}$ terms are determined individually
for the central arm and each muon arm based upon the detection of simulated 
$J/\psi$ processed using the real data analysis chain.  
Decay events are generated and propagated through a full 
GEANT simulation of the detector, which includes the specific details of 
the detector performance including the MuTr and MuID alignment,
disabled anodes and MuID efficiency.  
For the dielectron analysis, corrections to account for the detector dead 
channel map, energy calibration and run-to-run variations in the detector 
active area were determined from single electron yields.
The $J/\psi$ trigger efficiency is incorporated via a level-1 trigger 
emulator tuned to describe the experimental trigger response.
For the dimuon analysis, the level-2 filtering algorithms are applied 
to the simulated events.  
After reconstruction, the number of detected $J/\psi$ is compared to the 
number of simulated $J/\psi$ in a given rapidity and transverse momentum 
bin to determine the appropriate correction factors.

The systematic error associated with the measurement of the $J/\psi$ cross
section can be divided into three categories based upon the effect each error
has on the measured results.  All errors are reported as standard deviations.
Point-to-point uncorrelated errors, such as the
signal extraction systematic, which is bin dependent with typical values of 
$4\%$ ($5\%$) in the dimuon (dielectron) data, allow the
data points to move independently with respect to one another.  Point-to-point
correlated errors allow the data points to move coherently within the quoted
value.  Their values amount to $10\%$ ($8\%$) for the detector acceptance, 
$8\%$ ($4\%$) for the run-to-run variation in the detector efficiency, 
$4\%$ ($2.5\%$) for the $J/\psi$ transverse momentum and vertex distributions,  
and $2\%$ ($2\%$) for the hardware efficiency of the detector.  
Finally, global systematic errors allow the data points
to move together.  The dominant source of this error originates from the 
estimation of the BBC triggering efficiency for
minimum bias events, $9.7\%$, with additional contribution from the uncertainty 
in the estimation of the number of sampled
minimum bias events, $1\%$, and $\epsilon_{J/\psi}^{\rm BBC}$, $2.5\%$.

\begin{figure}[th]
\includegraphics[width=1.0\linewidth]{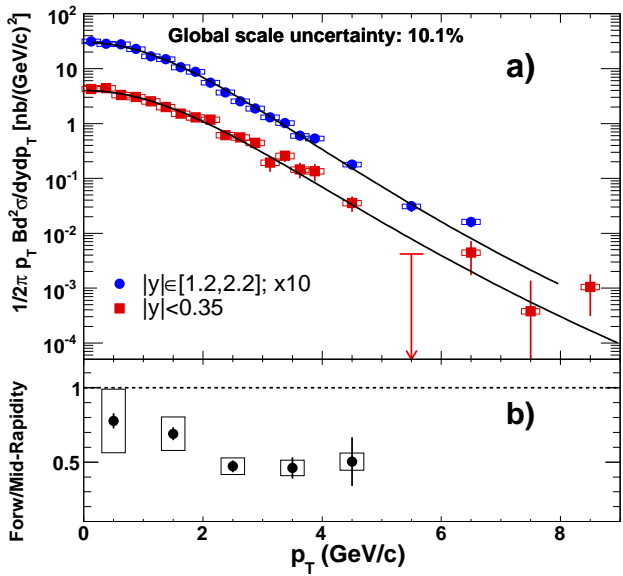}
\caption[]{(a) The mid and forward rapidity $J/\psi$ differential 
cross section times dilepton branching ratio versus $p_{T}$ and 
(b) the ratio of the mid and forward $p_{T}$ spectra.
The vertical error bars are the statistical and point-to-point 
uncorrelated error and the boxes are the point-to-point correlated 
systematic error. 
The solid lines are the fits described in the text.
\label{fig:cross_pt}
}
\end{figure}

Figure \ref{fig:cross_pt}(a) shows the transverse momentum spectra at both 
mid and forward rapidities, which are fit with the function, 
\mbox{A$\times(1+(p_{\rm T}/{\rm B})^2)^{-6}$}~\cite{Yoh}, to extract the 
$\langle p_{\rm T}^2\rangle$. 
At mid-rapidity 
$\langle p_{\rm T}^2\rangle = 4.14\pm0.18\pm^{0.30}_{0.20}$ (GeV/$c$)$^{2}$ 
and the $\chi^{2}$ per degree of freedom ($\chi^{2}/$ndf) is $23/19$.
At forward rapidity 
$\langle p_{\rm T}^2\rangle = 3.59\pm0.06\pm0.16$ (GeV/$c$)$^2$ 
and the $\chi^2/$ndf is $28/17$.
The first error is statistical and the second includes the systematic error 
from the maximum shape deviation permitted by the point-to-point 
correlated errors and from allowing the exponent of the fit function
to be a free parameter.
Although good agreement is found with the rapidity distribution and total 
cross section, previously published results \cite{ppg038} yielded a 
significantly lower $\langle p_{\rm T}^2\rangle$
at forward rapidity than found here, even accounting for the
quoted statistical and systematic uncertainties.  
The increased statistics of the present data set allow for an improved 
understanding of the shape of the $p_{\rm T}$ spectrum at forward rapidity due 
to the extended range in $p_{T}$ and the finer binning at low $p_{T}$.  
The previous results have been revisited and it was found that the systematic 
error was underestimated.

Figure~\ref{fig:cross_pt}(b) shows the ratio of the invariant cross 
section at forward and mid-rapidity.
The ratio falls with $p_{T}$ and reaches a minimum of 0.5 above a $p_{T}$ 
of 2 GeV$/c$.   
The data indicates that the forward rapidity $p_{T}$ distribution is 
substantially softer than mid-rapidity.
This is attributed to the increase in the longitudinal momentum at forward 
rapidity leaving less energy available in the transverse direction.

%%%%%%%%%%%%%%%%%%%%%%%%%%%%%%%%%%%%%%%%%%%%% Fig. 2

\begin{figure}[th]
\includegraphics[width=1.0\linewidth]{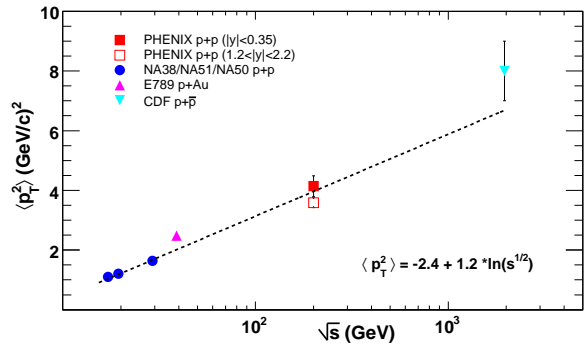}
\caption[]{PHENIX $\langle p_{\rm T}^2\rangle$ measurements compared to
measurements at other energies~\cite{E789,na51,CDFpt2} as a function of collision energy
for $J/\psi$ production in $p$+$p$ or $p$+$A$ collisions.
Also shown is a linear fit vs ln$(\sqrt{s})$.
\label{fig:pt2_roots}
}
\end{figure}

The observed $p_{\rm T}$ distributions are substantially harder than 
those for lower energy $p$+$p$ and $p$+$A$ collisions as expected from the 
increased phase space at higher energy.
Figure~\ref{fig:pt2_roots} shows the energy dependence of the average 
$\langle p_{\rm T}^2\rangle$ including lower energy points
from the Super Proton Synchrotron, and Fermilab fixed target and Tevatron 
measurements. 
A linear fit versus the log of the center of mass energy describes the general 
trends, although some variation is expected due to the differing rapidity 
ranges of the measurements and the use of $p+A$ data for some of the points.

%%%%%%%%%%%%%%%%%%%%%%%%%%%%%%%%%%%%%%%%%%%%% Fig. 3

%%%%%%%%%%%%%%%%%%%%%%%%%%%%%%%%%%%%%%%%%%%%% Fig. 4
\begin{figure}[hb]
\includegraphics[width=1.0\linewidth]{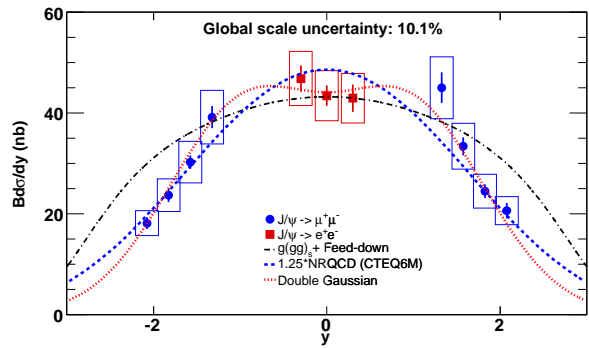}
\caption[]{The $J/\psi$ differential cross section times dilepton branching 
ratio plotted versus rapidity.  The vertical error bars represent the 
statistical and point-to-point uncorrelated error and the boxes represent 
the point-to-point correlated systematic error.
The curves are described in the text. \label{fig:cross_y}
}
\end{figure}

Figure~\ref{fig:cross_y} shows the $J/\psi$ differential cross section 
vs rapidity. The statistics of these results are large enough to allow 
eleven rapidity bins to be measured compared to 
the five bins used in the previous measurement~\cite{ppg038}.
Also shown are several model fits to the data. 
The dashed curve is a non-relativistic QCD calculation~\cite{nayak_nrqcd}.  
The dot-dash curve is a pQCD calculation that includes diagrams
describing a third gluon, which is necessary to neutralize the otherwise 
colored states~\cite{Khoze}.  
This model fails to reproduce the steeply falling cross 
section observed in the present data at large rapidity.
An empirical double Gaussian fit (dot-dot curve) is able to 
reproduce the data best, but has no strong theoretical foundation.
The data slightly favor a flatter distribution over the rapidity range 
$|y|<1.5$ than most models, 
but since the systematic error on the mid and forward rapidity points are 
independent, a narrower distribution is not excluded.

To determine the total cross section, the rapidity distribution was fit with 
many theoretical and phenomological shapes, including those shown in 
Fig.~\ref{fig:cross_y}. 
We obtain a total cross section times branching ratio of 
$B_{ll} \cdot \sigma_{pp}^{J/\psi} = 
178{\pm}3^{\rm stat}{\pm}53^{\rm sys}{\pm}18^{\rm norm}$~nb.
The absolute normalization error (norm) 
represents the uncertainty in the BBC trigger cross section.  
The systematic error (sys) is estimated from the maximum variation allowed 
by shifting the mid and forward rapidity data independently by their 
point-to-point correlated systematic errors.
This result is consistent with our previous measurement~\cite{ppg038}.

We have presented $J/\psi$ results for $p+p$ collisions 
at $\sqrt{s}$ = 200 GeV
that extend the reach in transverse momentum to 9 GeV/$c$.
The measured $p_{\rm T}$ spectrum is harder than that observed at lower energies 
and also shows a softening at forward rapidity.
The rapidity shape falls steeply at forward rapidity and can not be reproduced
by the pQCD calculation in~\cite{Khoze}.
Futhermore, the data slightly favor a flatter rapidity distribution than 
most models, but a narrower distribution is not excluded.
These data not only constrain production models for heavy quarkonia, but 
also provide a critical baseline for similar studies in deuteron-nucleus 
and heavy-ion collisions~\cite{ppg068,ppg071}.

%\input{15_Acknowledgements.tex}
%\section{Acknowledgements}   % Run-4 and Run-5 short form for PRL

We thank the staff of the Collider-Accelerator and
Physics Departments at BNL for their vital contributions.
We acknowledge support from
the Department of Energy and NSF (U.S.A.),
MEXT and JSPS (Japan),
CNPq and FAPESP (Brazil),
NSFC (China),
MSMT (Czech Republic),
IN2P3/CNRS, and CEA (France),
BMBF, DAAD, and AvH (Germany),
OTKA (Hungary),
DAE (India),
ISF (Israel),
KRF and KOSEF (Korea),
MES, RAS, and FAAE (Russia),
VR and KAW (Sweden),
U.S. CRDF for the FSU,
US-Hungarian NSF-OTKA-MTA,
and US-Israel BSF.


\begin{thebibliography}{37}
\expandafter\ifx\csname natexlab\endcsname\relax\def\natexlab#1{#1}\fi
\expandafter\ifx\csname bibnamefont\endcsname\relax
  \def\bibnamefont#1{#1}\fi
\expandafter\ifx\csname bibfnamefont\endcsname\relax
  \def\bibfnamefont#1{#1}\fi
\expandafter\ifx\csname citenamefont\endcsname\relax
  \def\citenamefont#1{#1}\fi
\expandafter\ifx\csname url\endcsname\relax
  \def\url#1{\texttt{#1}}\fi
\expandafter\ifx\csname urlprefix\endcsname\relax\def\urlprefix{URL }\fi
\providecommand{\bibinfo}[2]{#2}
\providecommand{\eprint}[2][]{\url{#2}}

\bibitem{beneke} M. Beneke, M. Kraemer, \Journal{\PRD} {55} {5269} {1997};
 M. Beneke, I. Z. Rothstein, \Journal{\PRD} {54} {2005} {1996}.

\bibitem{CDF} F.~Abe {\it et al.}, \Journal{\PRL} {69} {3704} {1992};
 F.~Abe {\it et al.}, \Journal{\PRD} {66} {092001} {2002};
 S.~Abachi {\it et al.}, \Journal{\PLB} {370} {239} {1996};
 B.~Abbot {\it et al.}, \Journal{\PRL} {82} {35} {1999}.

\bibitem{fixed_target} R.~Vogt, \Journal{\PLC} {310} {197} {1999}.

\bibitem{E789} M. H. Schub {\it et al.}, \Journal{\PRD} {52} {1307} {1995}.

\bibitem{HERAB} I. Abt {\it et al.}, \Journal{\PLB} {561} {61} {2003}. 

\bibitem{CSM}  E. L. Berger {\it et al.}, \Journal{\PRD} {23} {1521} {1981};
 R.~Baier {\it et al.}, \Journal{\PLB} {102} {364} {1981}.

\bibitem{COM} G.~T.~Bodwin {\it et al.}, \Journal{\PRD} {51} {1125} {1995};
 erratum \Journal{\PRD} {55} {5853} {1997}.

\bibitem{octet_matrix_elements} J. K. Mizukoshi, SLAC-PUB-8296, hep-ph/9911384 (1999).

\bibitem{Khoze} V. A. Khoze {\it et al.}, Eur. Phys. J. {\bf C39}, 163-171 (2005) and private communications.

\bibitem{ppg017}  S. S. Adler {\it et al.}, \Journal{\PRL} {92} {051802} {2004}.

\bibitem{ppg038}  S. S. Adler {\it et al.}, \Journal{\PRL} {96} {012304} {2006}.

\bibitem{phenix_nim} K. Adcox {\it et al.}, \Journal{\NIMA} {499} {469} {2003}.

\bibitem[{\citenamefont{Spiridonov}(2005)}]{radiative_effect} A. Spiridonov, hep-ex/0510076 (2004).

\bibitem{GEANT} GEANT 3.2.1, CERN Computing Library, \\ http://wwwasd.web.cern.ch/wwwasd/geant/index.html. 

\bibitem{Yoh}  J. K. Yoh {\it et al.}, \Journal{\PRL} {41} {684} {1978}; and private communication..

\bibitem{na51} O. Drapier, Th\`{e}se d'habilitation, Universit\'{e} Claude Bernard - Lyon 1, 1998; available at http://na50.web.cern.ch/NA50/theses.html.

\bibitem{CDFpt2} D. Acosta {\it et al.}, \Journal{\PRD} {71} {032001} {2005}.  The $\langle p_{\rm T}^2\rangle$ was extracted using a fit of the form $A\times(1+(p_{\rm T}/B)^2)^{-6}$.

\bibitem{nayak_nrqcd}  F. Cooper {\it et al.}, \Journal{\PRL} {93} {171801} {2004}; and private communication.

\bibitem{ppg068} S. S. Adler {\it et al.}, (Au+Au $J/\psi$ paper, preprint 
number soon).

\bibitem{ppg071} S. S. Adler {\it et al.}, (Cu+Cu $J/\psi$ paper, preprint 
number soon).

\end{thebibliography}
\end{document}